\documentclass[twocolumn,showpacs,preprintnumbers,amsmath,amssymb,aps,prb]{revtex4}

\usepackage{graphicx}
\usepackage{color}
\usepackage{longtable}
\begin{document}

\title{Spin correlations near the edge as probe of Dimer order \\in square-lattice Heisenberg models}
\author{T. Pardini}
\author{R.R.P. Singh}%
\email[R.R.P Singh: ]{singh@raman.physics.ucdavis.edu}
\affiliation{Department of Physics, University of California, Davis, California 95616,USA}
\date{Jan 7,2009}%
\begin{abstract}
Recent numerical and analytical work has shown that for the square-lattice
Heisenberg model the boundary can induce Dimer correlations near the edge
which are absent in spin-wave theories and non-linear sigma
model approaches. Here, we calculate the nearest-neighbor spin correlations
parallel and perpendicular to the boundary in a semi-infinite system
for two different square-lattice Heisenberg models: (i) A
frustrated $J_1-J_2$ model with nearest and second neighbor couplings and (ii)
a spatially anisotropic Heisenberg model, with nearest-neighbor
couplings $J$ perpendicular to the boundary and $J^\prime$
parallel to the boundary. We find that in the latter model, as $J^\prime/J$
is reduced from unity the Dimer correlations near the edge become longer
ranged. In contrast, in the frustrated model, with increasing $J_2$,
dimer correlations are
strengthened near the boundary but they decrease rapidly with
distance. These results imply that deep inside the N\'{e}el phase of the $J_1-J_2$
Heisenberg model, dimer correlations remain short-ranged.
Hence, if there is a direct transition between the two it is either
first order or there is a very narrow critical region.
\end{abstract}
\maketitle
\section{Introduction}
Square-lattice antiferromagnets have been studied extensively
in recent years~\cite{CHN,Read1989,Gelfand1989,Read1991,Oitmaa1996,Sandvik1997,Singh1999,Capriotti2000,Zheng2005}. Yet, new surprises continue to arise. In particular,
recent Quantum Monte Carlo studies~\cite{Hoglund2008} by H\"{o}glund and Sandvik have shown that the existence
of a free edge induces pronounced Dimerized correlations in the system. 
In a follow up work~\cite{Metlitski2008} it was shown by Metlitski and Sachdev that the presence of a 
boundary induces dimer
correlations perpendicular to the boundary. And since the correlations
decay with distance from the boundary, their gradient induces
alternation in the spin-correlations parallel
to the boundary, leading to specific pattern of nearest-neighbor
spin correlations observed by H\"{o}glund and Sandvik in their simulations.
These effects are absent in spin-wave theories and
non-linear sigma model approaches. 

Over the past few years there has been considerable interest in
the possibility of direct continuous phase transitions between N\'{e}el
and Valence Bond Crystal (VBC) phases~\cite{Schulz1996,Senthil2004,Senthil2004a,Kulov2006,Isakov2006,Melko2006,Damle2006,Kulov2004,Kragset2006,Nogueira2007}. Such phase transitions have
been called deconfined quantum criticality and are marked by the
liberation of spin-half degrees of freedom as well as the existence
of massless spin-singlet photon field. Strong numerical evidence
for such a scenario has been provided in Sandvik's J-Q model~\cite{Sandvik2007,Kaul2008,Melko2008},
where the Heisenberg model is supplemented by a 4-spin interaction
around a plaquette. An alternative possibility of a weakly first
order transition has also been raised~\cite{Kuklov2008,Jiang2008}.

A more realistic model of two-dimensional square-lattice 
quantum antiferromagnets is the spin-half
$J_1-J_2$ model, where there is nearest-neighbor interaction $J_1$ and
second neighbor interaction $J_2$. Increasing $J_2$ increases spin
frustration and is known to lead to a magnetically disordered state
at intermediate $J_2/J_1$ values~\cite{Oitmaa1996,Singh1999}. There is substantial and growing
body of numerical evidence that
the magnetically disordered phase has Valence Bond Crystal (VBC) order~\cite{Read1989,Gelfand1989,Schulz1996,Singh1999}. The question of
whether the transition between the N\'{e}el and Dimer orders is continuous
or first order remains a subject of debate~\cite{Sirker2006,Darradi2008}.

Here, we would like to use the edge induced dimer correlations as
a probe of growth of dimer correlations inside the N\'{e}el phase and
thus address the possibility of a diverging dimer correlation length
in the N\'{e}el phase. We study two models. A spatially anisotropic
model with interactions $J$ and $J^\prime$ along the two axes. We
choose the boundary to be parallel to the direction of the weaker
coupling $J^\prime$. It is well known that one-dimensional Heisenberg
model has power-law decaying Valence Bond correlations. Thus as
one approaches the limit of small $J^\prime$ one expects to see
the edge induced correlations to have a long length scale. This model acts
as a test case for our method. We also study the $J_1-J_2$
Heisenberg model. It is for this model that one would like to
see how the range of dimer correlations grows near the boundary
as spin-frustration given by the parameter $J_2/J_1$ increases
and one approaches the phase transition, where N\'{e}el order is lost.

\section{Series expansion}\label{seriesexpansion}
\begin{figure}[!h]\label{Fig1}
\resizebox{75mm}{!}{\includegraphics{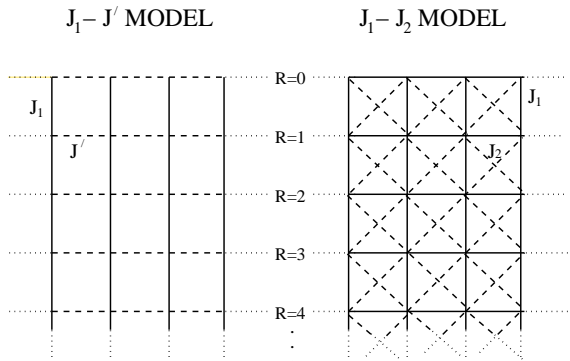}}
\caption{\label{fig:Fig1} (left) $J_1-J^{\prime}$ model on the semi-infinite square lattice. The interaction $J_1=1$ is perpendicular to the edge, 
the interaction $J^{\prime}$ is parallel to the edge and it is allowed to vary in the range $0<J^{\prime}\leq 1$. (Right) $J_1-J_2$ model 
on the semi-infinite square lattice. $J_1=1$ and $0\leq J_2<0.3$.
The parameter $R$ defines the distance of a particular lattice plane from the edge.}
\end{figure}

The antiferromagnetic Heisenberg models defined by two coupling constants $J_1$ and $J_2$
(or by $J$ and $J^\prime$) are shown in Fig.~\ref{fig:Fig1}. We consider a semi-infinite system, with
a boundary parallel to the $X$-axis
also shown in Fig.~1. Since we are considering a system inside a colinear N\'{e}el ordered
phase, we develop an Ising series expansion~\cite{Gelfand2000}, where all Heisenberg couplings are
written as 
\begin{equation}
\textbf{S}_i\cdot \textbf{S}_j=S_i^z S_j^z +\lambda (S_i^x S_j^x +S_i^y S_j^y).
\end{equation}
The parameter $\lambda$ acts as an expansion parameter. We develop series expansions
for on-site local magnetization $\langle S_i^z \rangle$ as well as for nearest neighbor spin-correlations
$\langle \textbf{S}_i\cdot \textbf{S}_j \rangle$, parallel and perpendicular to the boundary. In the
semi-infinite system, these quantities depend on the distance $R$ from the boundary.
In the series expansion method, the boundary can be accomodated by accounting for the
graphs that terminate at the boundary. Apart from this, the formalism of linked cluster
expansions remains unchanged.

\section{Results: Correlations and Excitations near the Edge}
\begin{figure}[!h]
  \resizebox{85mm}{!}{\includegraphics{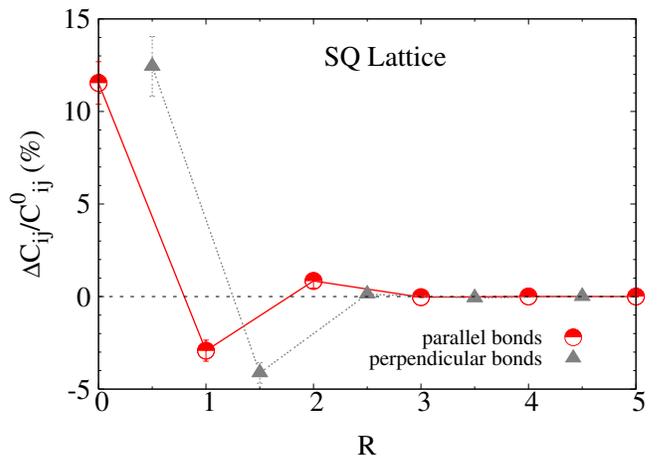}}
  \caption{\label{fig:Fig2}(Color online) Correlation function of the spin-$\frac{1}{2}$ Heisenberg model on the semi-infinite (SI) 
    square lattice for bonds parallel and perpendicular to the edge as a function of distance R. Notice that bonds perpendicular to the edge 
have their centers at half-integer values of $R$.
The $y$-axis is $\Delta C_{ij}/C^0_{ij}=-\frac{\langle\vec{S_i}\cdot \vec{S_j}\rangle_{SI}-\langle\vec{S_i}\cdot \vec{S_j}\rangle_{\infty}}{\langle\vec{S_i}\cdot \vec{S_j}\rangle_{\infty}}$, where the $\infty$ index refers to the bulk value.}
\end{figure}

First we present the results for the square-lattice Heisenberg model.
The nearest neighbor correlations parallel and perpendicular to the boundary
are shown in Fig.~\ref{fig:Fig2}. These are obtained by d-log Pad\'{e} approximant analysis of the series. 
They agree well with the results of Sandvik and H\"{o}glund~\cite{Hoglund2008}. The important thing to note is that 
they both decrease rapidly with distance and by $R=5$ they differ from the bulk value by less than $0.1\%$.

\begin{figure}[!h]
  \resizebox{80mm}{!}{\includegraphics{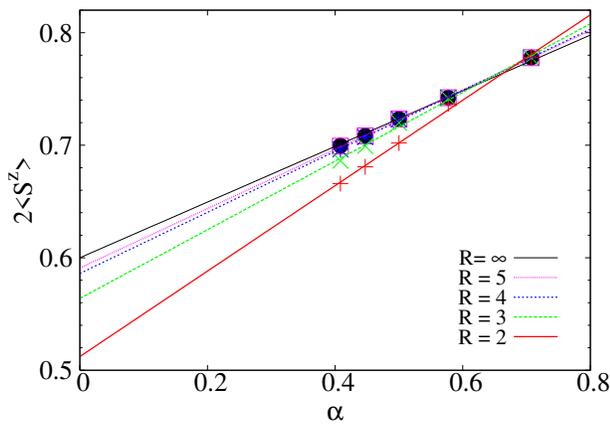}}
  \caption{\label{fig:Fig3}(Color online). Partial sums of series expansion coefficients for the on-site sublattice magnetization of the semi-infinte 
    square lattice model. The fit for different values of the parameter $R$ are shown. See text for details}
\end{figure}

The on-site magnetization also changes near the boundary. The results for magnetization are more 
sensitive to extrapolation methods than spin-spin correlations because 
one expects a square root singularity for this quantity. This means that contributions of higher order
terms only decay as $1/\sqrt{N}$.
We have followed the method used in Ref.~\onlinecite{Singh1989} for the bulk system.
We obtain partial sums $S_N$ of series coefficients and then fit them vs $\alpha =\frac{1}{\sqrt{1+N}}$ to estimate $S_N$ as $N\rightarrow \infty$. These are shown 
in Fig.~\ref{fig:Fig3} for values of $R\geq 2$. We deduce the uncertainty in the magnetization by the uncertainty
in the linear fits.

\begin{figure}[!h]
  \resizebox{75mm}{!}{\includegraphics{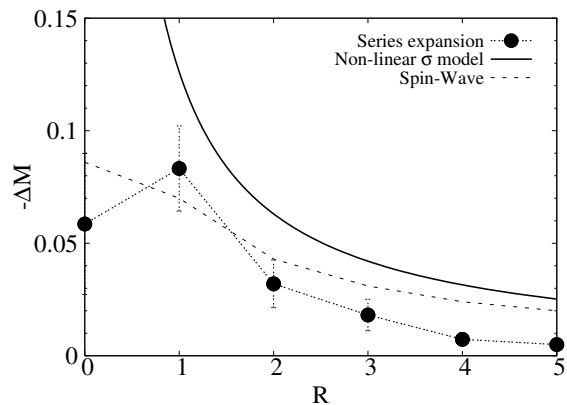}}
  \caption{\label{fig:Fig4} On-site sublattice magnetization for the spin-$\frac{1}{2}$ Heisenberg model on the semi-infinite square lattice. The non-linear sigma model and 
    spin wave results from Ref.~\onlinecite{Metlitski2008} are also shown. The $y$-axis is $\Delta M=(M_{SI}-M_{\infty})$.}
\end{figure}

Results obtained this way are plotted in Fig.~\ref{fig:Fig4} where they are compared to the non-linear $\sigma$ model 
and spin wave results~\cite{Metlitski2008}. The on-site sublattice magnetization is diminished at the edge and its reduction is comparable to what is 
obtained in spin-wave theory. Away from the edge the sublattice magnetization should approach its bulk value. In the non-linear $\sigma$ model and spin wave theory
the change in magnetization follows a $1/R$ behavior. 
On general grounds one expects
the non-linear $\sigma$ model results, when expressed in terms of renormalized
parameters, to be exact\cite{Metlitski2008} for large-R. 
The reduction is less in our calculation upto the largest distance studied, that is,
$R=5$. 
Part of the reason maybe that the
asymptotic behavior may set in at significantly large-R due to the dimer-correlations
at the boundary. However, it is also likely that the uncertainty in our calculations
are much larger than shown. Our estimate of the bulk magnetization is $0.302$. If we
replace it by the more accurate results from higher order series expansions~\cite{Zheng2005}
or quantum Monte Carlo simulations~\cite{Sandvik2007}, which is $0.307$ it would shift our
calculated curves up by $0.005$ and bring them closer to the spin-wave results. This discrepency
in the bulk estimates 
implies that the uncertainties are much bigger than estimated by the fits and they are
particularly magnified at larger $R$ because we are taking the difference of two quantities which are close
in magnitude.

\begin{figure}
  \resizebox{80mm}{!}{\includegraphics{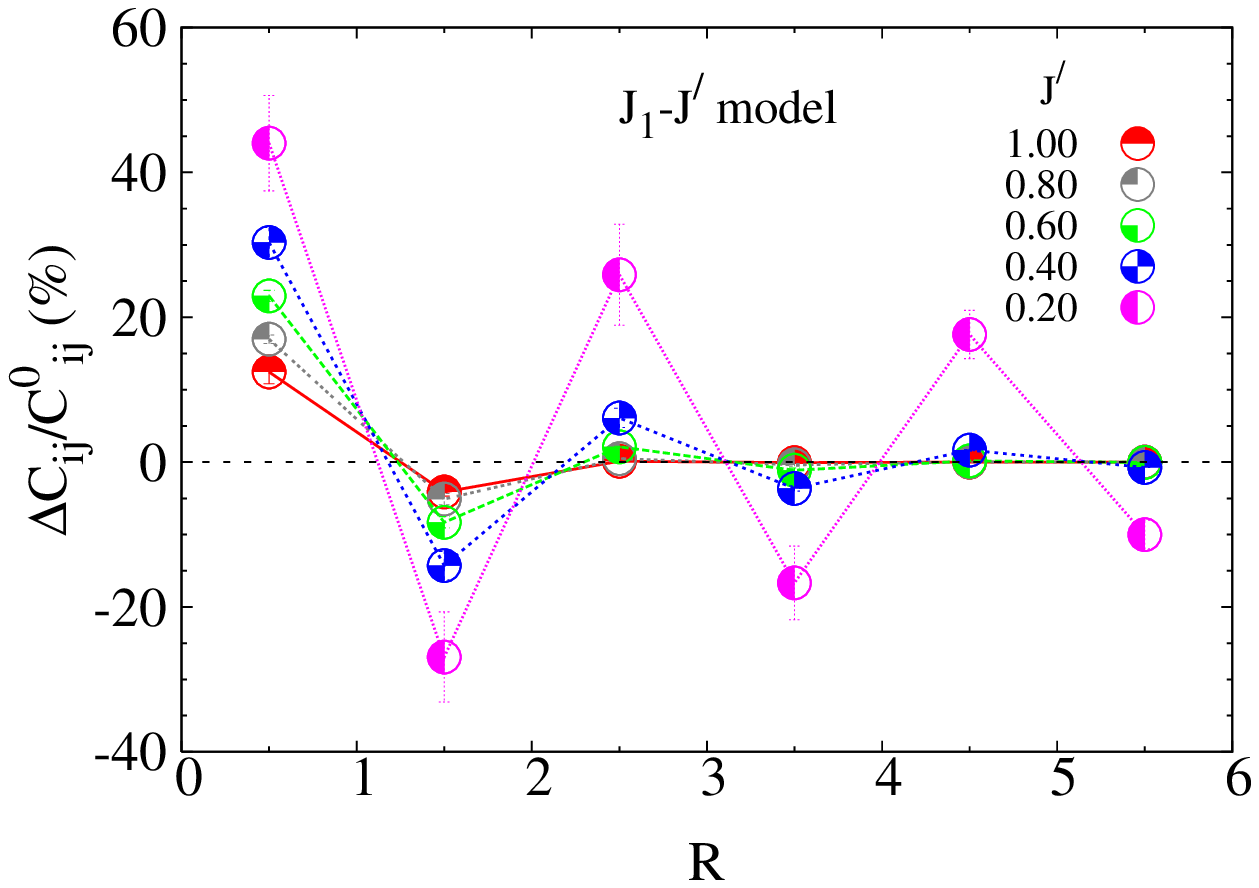}}
  \caption{\label{fig:Fig5} (Color online) Correlation function of the spin-$\frac{1}{2}$ $J_1-J^{\prime}$ model on the semi-infinite 
    square lattice for bonds perpendicular to the edge as a function of distance R for selected values of $J^{\prime}$. The quantity  $\Delta C_{ij}/C^0_{ij}$ shown on the 
    $y$-axis is defined in the caption of Fig~\ref{fig:Fig2}.}
\end{figure}

In Fig.~\ref{fig:Fig5}, we show the nearest-neighbor spin-correlations perpendicular
to the boundary for the $J-J^\prime$ model. This is the direction of the stronger coupling.
In the one-d limit, one expects the free end to induce dimer correlations in the system
that decay as a power-law away from the boundary. Indeed, we find that as the system becomes
more and more anisotropic, the dimer correlations become more and more long ranged.

\begin{figure}
  \resizebox{80mm}{!}{\includegraphics{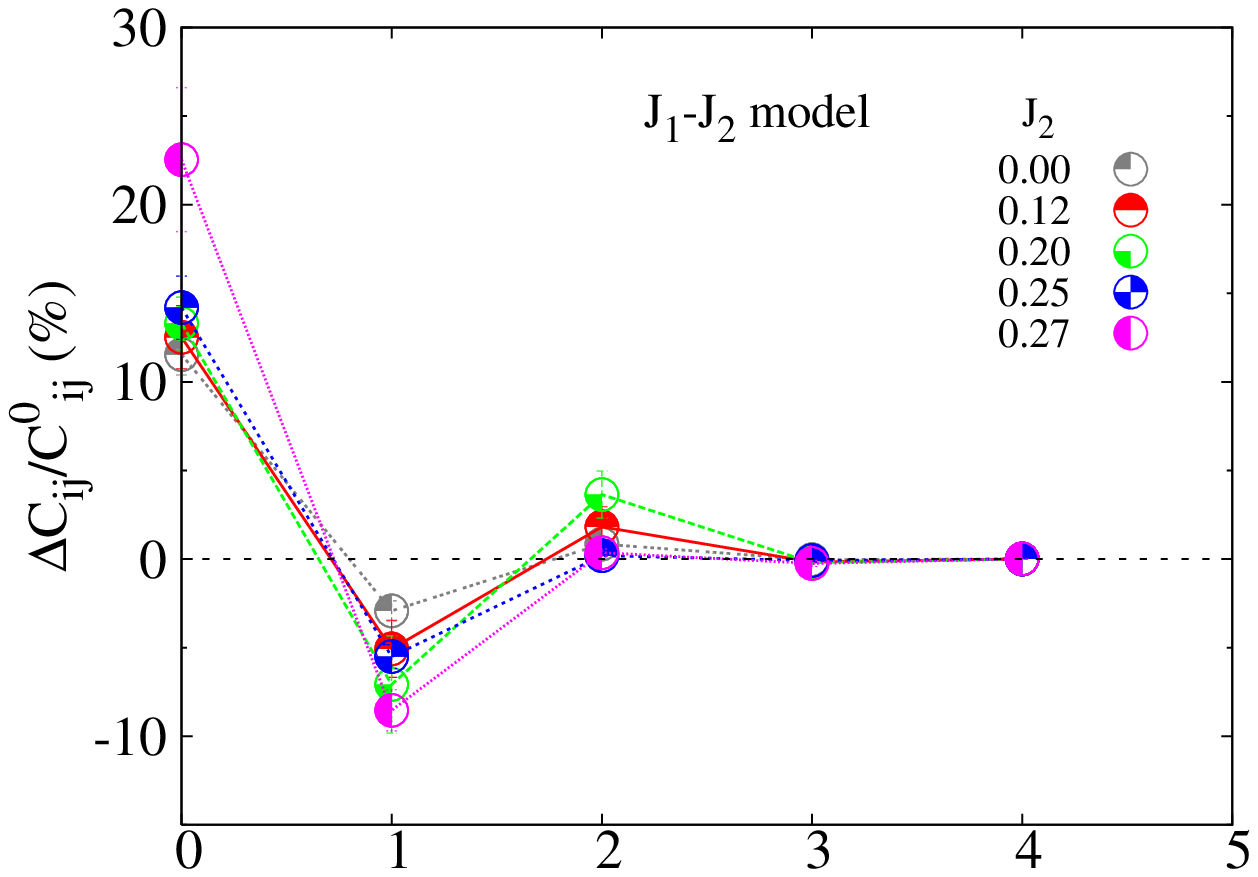}}
  \resizebox{80mm}{!}{\includegraphics{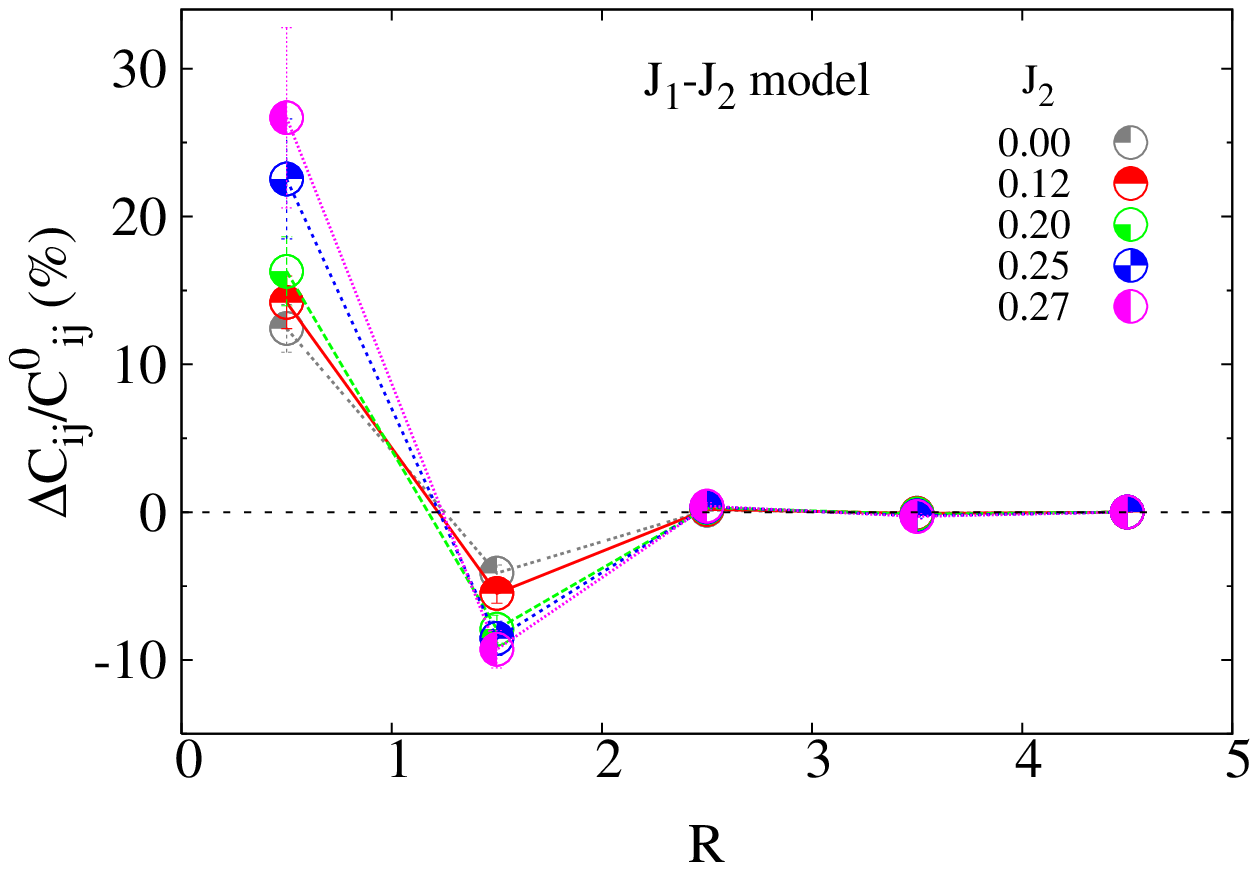}}
\caption{\label{fig:Fig6} (Color online) Correlation function of the spin-$\frac{1}{2}$ $J_1-J_2$ model on the semi-infinite 
  square lattice for bonds parallel (top panel) and perpendicular (bottom panel) to the edge as a function of distance R for selected values of $J_2$. 
  The quantity  $\Delta C_{ij}/C^0_{ij}$ shown on the $y$-axis is defined in the caption of Fig~\ref{fig:Fig2}. }
\end{figure}
In Fig.~\ref{fig:Fig6}, we show the nearest-neighbor spin-correlations for the $J_1-J_2$ model.
Correlations both parallel and perpendicular to the boundary are shown. In this case,
we find that while a frustrating second neighbor interaction enhances the dimerization
near the boundary, it does not appear to increase the range over which dimer correlations
extend. The convergence of our analysis becomes poor as we get close to the bulk transition
away from N\'{e}el order, which has been estimated to be in the range $J_2/J_1\approx 0.35-0.4$.\cite{Oitmaa1996,Singh1999,Sirker2006}

These results show that in the $J_1-J_2$ square-lattice Heisenberg model, one does not
have appreciable range Valence Bond Correlation in the bulk even with significant frustration.
They suggest that a direct transition between N\'{e}el and Dimer phases is likely 
first order. Our study
can not rule out the possibility that the dimer correlations build up very quickly close to
the transition. This would imply  a very narrow critical region in this model.
\begin{figure}[!ht]
\resizebox{70mm}{!}{\includegraphics{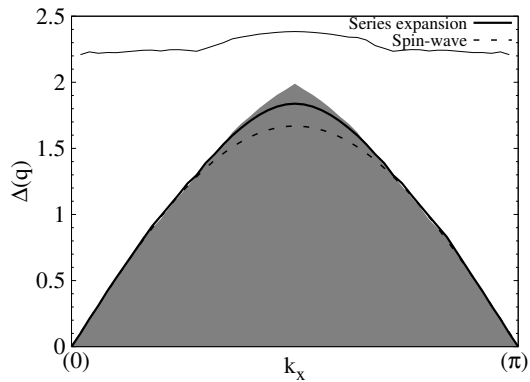}}
\caption{\label{fig:Fig7} Excitation spectrum of the Heisenberg model on the semi-infinite square lattice 
for $0\leq k_x\leq \pi$, where $k_x$ is the momentum parallel to the edge. 
The bound states found by spin-wave theory and series expansion calculations are shown. The thin solid line represents the upper limit of the continuum of states.}
\end{figure}

We have also calculated the spin-wave spectrum for the magnon states that are bound to
the surface for the nearest-neighbor square-lattice Heisenberg model. 
The momentum parallel to the surface is a good quantum number. In the
series expansion calculations, the spin-flip states right at the boundary have a different excitation energy
from those which are away from the boundary. Thus these states get separated from the bulk
states starting in zeroth order. Upon extrapolation to the Heisenberg model, we find
the dispersion of these surface magnons as shown in the Figure~\ref{fig:Fig7}. Also, shown are
results from the spin-wave calculations of Metlitski and Sachdev~\cite{Metlitski2008}. The latter has been
renormalized to have the same spin-wave velocity as the bulk. Our results are
in agreement with the latter that for a large part of the Brillouin zone, the
surface states are hugging the continuum. Only very near $k=\pi/2$, they clearly
separate from the continuum. In this region, the binding energy in our
calculation is smaller than in spin-wave theory.

\section{Conclusions}
In this paper we have studied the spin-correlations and excitations near
the boundary of two dimensional Heisenberg antiferromagnets. Two different
square-lattice models are considered. One where the exchange coupling 
parallel to the boundary is smaller than those perpendicular to the
boundary. In this model, we find that the boundary induced dimerization
becomes more and more long-ranged as the anisotropy is increased.
The second model is the $J_1-J_2$ Heisenberg model, with nearest and
second neighbor exchange interactions. In this case, we find that
as frustration is increased in the model, the boundary induced
dimerization increases close to the boundary but its range does
not change significantly. This suggests that in the $J_1-J_2$ model,
the N\'{e}el phase does not develop long-range dimer correlations. Hence,
either the transition from N\'{e}el to dimer order is first order in
this model, or
there is a very narrow critical region.

Acknowledgements: We would like to thank Subir Sachdev and Max Metlitski for
very useful discussions.


\begin{thebibliography}{30}
\expandafter\ifx\csname natexlab\endcsname\relax\def\natexlab#1{#1}\fi
\expandafter\ifx\csname bibnamefont\endcsname\relax
  \def\bibnamefont#1{#1}\fi
\expandafter\ifx\csname bibfnamefont\endcsname\relax
  \def\bibfnamefont#1{#1}\fi
\expandafter\ifx\csname citenamefont\endcsname\relax
  \def\citenamefont#1{#1}\fi
\expandafter\ifx\csname url\endcsname\relax
  \def\url#1{\texttt{#1}}\fi
\expandafter\ifx\csname urlprefix\endcsname\relax\def\urlprefix{URL }\fi
\providecommand{\bibinfo}[2]{#2}
\providecommand{\eprint}[2][]{\url{#2}}

\bibitem[{\citenamefont{Chakravarty et~al.}(1989)\citenamefont{Chakravarty,
  Halperin, and Nelson}}]{CHN}
\bibinfo{author}{\bibfnamefont{S.}~\bibnamefont{Chakravarty}},
  \bibinfo{author}{\bibfnamefont{B.~I.} \bibnamefont{Halperin}},
  \bibnamefont{and} \bibinfo{author}{\bibfnamefont{D.~R.}
  \bibnamefont{Nelson}}, \bibinfo{journal}{Phys. Rev. B.}
  \textbf{\bibinfo{volume}{39}}, \bibinfo{pages}{2344} (\bibinfo{year}{1989}).

\bibitem[{\citenamefont{Read and Sachdev}(1989)}]{Read1989}
\bibinfo{author}{\bibfnamefont{N.}~\bibnamefont{Read}} \bibnamefont{and}
  \bibinfo{author}{\bibfnamefont{S.}~\bibnamefont{Sachdev}},
  \bibinfo{journal}{Phys. Rev. Lett.} \textbf{\bibinfo{volume}{62}},
  \bibinfo{pages}{1694} (\bibinfo{year}{1989}).

\bibitem[{\citenamefont{Gelfand et~al.}(1989)\citenamefont{Gelfand, Singh, and
  Huse}}]{Gelfand1989}
\bibinfo{author}{\bibfnamefont{M.~P.} \bibnamefont{Gelfand}},
  \bibinfo{author}{\bibfnamefont{R.~R.~P.} \bibnamefont{Singh}},
  \bibnamefont{and} \bibinfo{author}{\bibfnamefont{D.~A.} \bibnamefont{Huse}},
  \bibinfo{journal}{Phys. Rev. B} \textbf{\bibinfo{volume}{40}},
  \bibinfo{pages}{10801} (\bibinfo{year}{1989}).

\bibitem[{\citenamefont{Read and Sachdev}(1991)}]{Read1991}
\bibinfo{author}{\bibfnamefont{N.}~\bibnamefont{Read}} \bibnamefont{and}
  \bibinfo{author}{\bibfnamefont{S.}~\bibnamefont{Sachdev}},
  \bibinfo{journal}{Phys. Rev. Lett.} \textbf{\bibinfo{volume}{66}},
  \bibinfo{pages}{1773} (\bibinfo{year}{1991}).

\bibitem[{\citenamefont{Oitmaa and Zheng}(1996)}]{Oitmaa1996}
\bibinfo{author}{\bibfnamefont{J.}~\bibnamefont{Oitmaa}} \bibnamefont{and}
  \bibinfo{author}{\bibfnamefont{W.}~ \bibnamefont{Zheng}},
  \bibinfo{journal}{Phys. Rev. B.} \textbf{\bibinfo{volume}{54}},
  \bibinfo{pages}{3022} (\bibinfo{year}{1996}).

\bibitem[{\citenamefont{Sandvik}(1997)}]{Sandvik1997}
\bibinfo{author}{\bibfnamefont{A.~W.} \bibnamefont{Sandvik}},
  \bibinfo{journal}{Phys. Rev. B.} \textbf{\bibinfo{volume}{56}},
  \bibinfo{pages}{11678} (\bibinfo{year}{1997}).

\bibitem[{\citenamefont{Singh et~al.}(1999)\citenamefont{Singh, Weihong, Hamer,
  and Oitmaa}}]{Singh1999}
\bibinfo{author}{\bibfnamefont{R.~R.~P.} \bibnamefont{Singh}},
  \bibinfo{author}{\bibfnamefont{Z.}~\bibnamefont{Weihong}},
  \bibinfo{author}{\bibfnamefont{C.~J.} \bibnamefont{Hamer}}, \bibnamefont{and}
  \bibinfo{author}{\bibfnamefont{J.}~\bibnamefont{Oitmaa}},
  \bibinfo{journal}{Phys. Rev. B} \textbf{\bibinfo{volume}{60}},
  \bibinfo{pages}{7278} (\bibinfo{year}{1999}).

\bibitem[{\citenamefont{Capriotti and Sorella}(2000)}]{Capriotti2000}
\bibinfo{author}{\bibfnamefont{L.}~\bibnamefont{Capriotti}} \bibnamefont{and}
  \bibinfo{author}{\bibfnamefont{S.}~\bibnamefont{Sorella}},
  \bibinfo{journal}{Phys. Rev. Lett.} \textbf{\bibinfo{volume}{84}},
  \bibinfo{pages}{3173} (\bibinfo{year}{2000}).

\bibitem[{\citenamefont{Zheng et~al.}(2005)\citenamefont{Zheng, Oitmaa, and
  Hamer}}]{Zheng2005}
\bibinfo{author}{\bibfnamefont{W.}~\bibnamefont{Zheng}},
  \bibinfo{author}{\bibfnamefont{J.}~\bibnamefont{Oitmaa}}, \bibnamefont{and}
  \bibinfo{author}{\bibfnamefont{C.~J.} \bibnamefont{Hamer}},
  \bibinfo{journal}{Phys. Rev. B.} \textbf{\bibinfo{volume}{71}},
  \bibinfo{pages}{184440} (\bibinfo{year}{2005}).

\bibitem[{\citenamefont{H\"{o}glund and Sandvik}()}]{Hoglund2008}
\bibinfo{author}{\bibfnamefont{K.~H.} \bibnamefont{H\"{o}glund}}
  \bibnamefont{and} \bibinfo{author}{\bibfnamefont{A.~W.}
  \bibnamefont{Sandvik}}, \bibinfo{howpublished}{arXiv:0808.0408v1}.

\bibitem[{\citenamefont{Metlitski and Sachdev}(2008)}]{Metlitski2008}
\bibinfo{author}{\bibfnamefont{M.~A.} \bibnamefont{Metlitski}}
  \bibnamefont{and} \bibinfo{author}{\bibfnamefont{S.}~\bibnamefont{Sachdev}},
  \bibinfo{journal}{Phys. Rev. B.} \textbf{\bibinfo{volume}{78}},
  \bibinfo{pages}{174410} (\bibinfo{year}{2008}).

\bibitem[{\citenamefont{Schulz et~al.}(1996)\citenamefont{Schulz, Ziman, and
  Poilblanc}}]{Schulz1996}
\bibinfo{author}{\bibfnamefont{H.~J.} \bibnamefont{Schulz}},
  \bibinfo{author}{\bibfnamefont{T.}~\bibnamefont{Ziman}}, \bibnamefont{and}
  \bibinfo{author}{\bibfnamefont{J.}~\bibnamefont{Poilblanc}},
  \bibinfo{journal}{J. de Phys. I} \textbf{\bibinfo{volume}{6}},
  \bibinfo{pages}{675} (\bibinfo{year}{1996}).

\bibitem[{\citenamefont{Senthil
  et~al.}(2004{\natexlab{a}})\citenamefont{Senthil, Vishwanath, Balents,
  Sachdev, and Fisher}}]{Senthil2004}
\bibinfo{author}{\bibfnamefont{T.}~\bibnamefont{Senthil}},
  \bibinfo{author}{\bibfnamefont{A.}~\bibnamefont{Vishwanath}},
  \bibinfo{author}{\bibfnamefont{L.}~\bibnamefont{Balents}},
  \bibinfo{author}{\bibfnamefont{S.}~\bibnamefont{Sachdev}}, \bibnamefont{and}
  \bibinfo{author}{\bibfnamefont{M.}~\bibnamefont{Fisher}},
  \bibinfo{journal}{Science} \textbf{\bibinfo{volume}{303}},
  \bibinfo{pages}{1490} (\bibinfo{year}{2004}{\natexlab{a}}).

\bibitem[{\citenamefont{Senthil
  et~al.}(2004{\natexlab{b}})\citenamefont{Senthil, Balents, Sachdev,
  Vishwanath, and Fisher}}]{Senthil2004a}
\bibinfo{author}{\bibfnamefont{T.}~\bibnamefont{Senthil}},
  \bibinfo{author}{\bibfnamefont{L.}~\bibnamefont{Balents}},
  \bibinfo{author}{\bibfnamefont{S.}~\bibnamefont{Sachdev}},
  \bibinfo{author}{\bibfnamefont{A.}~\bibnamefont{Vishwanath}},
  \bibnamefont{and} \bibinfo{author}{\bibfnamefont{M.}~\bibnamefont{Fisher}},
  \bibinfo{journal}{Phys. Rev. B.} \textbf{\bibinfo{volume}{70}},
  \bibinfo{pages}{144407} (\bibinfo{year}{2004}{\natexlab{b}}).

\bibitem[{\citenamefont{Kuklov et~al.}(2006)\citenamefont{Kuklov, Prokof'ev,
  Svistunov, and Troyer}}]{Kulov2006}
\bibinfo{author}{\bibfnamefont{A.}~\bibnamefont{Kuklov}},
  \bibinfo{author}{\bibfnamefont{N.}~\bibnamefont{Prokof'ev}},
  \bibinfo{author}{\bibfnamefont{B.}~\bibnamefont{Svistunov}},
  \bibnamefont{and} \bibinfo{author}{\bibfnamefont{M.}~\bibnamefont{Troyer}},
  \bibinfo{journal}{Ann. Phys. (N.Y.)} \textbf{\bibinfo{volume}{321}},
  \bibinfo{pages}{1602} (\bibinfo{year}{2006}).

\bibitem[{\citenamefont{Isakov et~al.}(2006)\citenamefont{Isakov, Wessel,
  Melko, Sengupta, and Kim}}]{Isakov2006}
\bibinfo{author}{\bibfnamefont{S.~V.} \bibnamefont{Isakov}},
  \bibinfo{author}{\bibfnamefont{S.}~\bibnamefont{Wessel}},
  \bibinfo{author}{\bibfnamefont{R.~G.} \bibnamefont{Melko}},
  \bibinfo{author}{\bibfnamefont{K.}~\bibnamefont{Sengupta}}, \bibnamefont{and}
  \bibinfo{author}{\bibfnamefont{Y.~B.} \bibnamefont{Kim}},
  \bibinfo{journal}{Phys. Rev. Lett.} \textbf{\bibinfo{volume}{97}},
  \bibinfo{pages}{147202} (\bibinfo{year}{2006}).

\bibitem[{\citenamefont{Melko et~al.}(2006)\citenamefont{Melko, Maestro, and
  Burkov}}]{Melko2006}
\bibinfo{author}{\bibfnamefont{R.~G.} \bibnamefont{Melko}},
  \bibinfo{author}{\bibfnamefont{A.~D.} \bibnamefont{Maestro}},
  \bibnamefont{and} \bibinfo{author}{\bibfnamefont{A.~A.}
  \bibnamefont{Burkov}}, \bibinfo{journal}{Phys. Rev. B.}
  \textbf{\bibinfo{volume}{74}}, \bibinfo{pages}{214517}
  (\bibinfo{year}{2006}).

\bibitem[{\citenamefont{Damle and Senthil}(2006)}]{Damle2006}
\bibinfo{author}{\bibfnamefont{K.}~\bibnamefont{Damle}} \bibnamefont{and}
  \bibinfo{author}{\bibfnamefont{T.}~\bibnamefont{Senthil}},
  \bibinfo{journal}{Phys. Rev. Lett.} \textbf{\bibinfo{volume}{97}},
  \bibinfo{pages}{067202} (\bibinfo{year}{2006}).

\bibitem[{\citenamefont{Kuklov et~al.}(2004)\citenamefont{Kuklov, Prokof'ev,
  and Svistunov}}]{Kulov2004}
\bibinfo{author}{\bibfnamefont{A.}~\bibnamefont{Kuklov}},
  \bibinfo{author}{\bibfnamefont{N.}~\bibnamefont{Prokof'ev}},
  \bibnamefont{and}
  \bibinfo{author}{\bibfnamefont{B.}~\bibnamefont{Svistunov}},
  \bibinfo{journal}{Phys. Rev. Lett.} \textbf{\bibinfo{volume}{93}},
  \bibinfo{pages}{230402} (\bibinfo{year}{2004}).

\bibitem[{\citenamefont{Kragset et~al.}(2006)\citenamefont{Kragset,
  Sm{\o}rgrav, Hove, Nogueira, and A.Sudb\o}}]{Kragset2006}
\bibinfo{author}{\bibfnamefont{S.}~\bibnamefont{Kragset}},
  \bibinfo{author}{\bibfnamefont{E.}~\bibnamefont{Sm{\o}rgrav}},
  \bibinfo{author}{\bibfnamefont{J.}~\bibnamefont{Hove}},
  \bibinfo{author}{\bibfnamefont{F.~S.} \bibnamefont{Nogueira}},
  \bibnamefont{and} \bibinfo{author}{\bibnamefont{A.Sudb\o}},
  \bibinfo{journal}{Phys. Rev. Lett.} \textbf{\bibinfo{volume}{97}},
  \bibinfo{pages}{247201} (\bibinfo{year}{2006}).

\bibitem[{\citenamefont{Nogueira et~al.}(2007)\citenamefont{Nogueira, Kragset,
  and Sudb\o}}]{Nogueira2007}
\bibinfo{author}{\bibfnamefont{F.~S.} \bibnamefont{Nogueira}},
  \bibinfo{author}{\bibfnamefont{S.}~\bibnamefont{Kragset}}, \bibnamefont{and}
  \bibinfo{author}{\bibfnamefont{A.}~\bibnamefont{Sudb\o}},
  \bibinfo{journal}{Phys. Rev. B.} \textbf{\bibinfo{volume}{76}},
  \bibinfo{pages}{220403} (\bibinfo{year}{2007}).

\bibitem[{\citenamefont{Sandvik}(2007)}]{Sandvik2007}
\bibinfo{author}{\bibfnamefont{A.~W.} \bibnamefont{Sandvik}},
  \bibinfo{journal}{Phys. Rev. Lett.} \textbf{\bibinfo{volume}{98}},
  \bibinfo{pages}{227202} (\bibinfo{year}{2007}).

\bibitem[{\citenamefont{Kaul and Melko}(2008)}]{Kaul2008}
\bibinfo{author}{\bibfnamefont{R.~K.} \bibnamefont{Kaul}} \bibnamefont{and}
  \bibinfo{author}{\bibfnamefont{R.~G.} \bibnamefont{Melko}},
  \bibinfo{journal}{Phys. Rev. B.} \textbf{\bibinfo{volume}{78}},
  \bibinfo{pages}{014417} (\bibinfo{year}{2008}).

\bibitem[{\citenamefont{Melko and Kaul}(2008)}]{Melko2008}
\bibinfo{author}{\bibfnamefont{R.~G.} \bibnamefont{Melko}} \bibnamefont{and}
  \bibinfo{author}{\bibfnamefont{R.~K.} \bibnamefont{Kaul}},
  \bibinfo{journal}{Phys. Rev. Lett.} \textbf{\bibinfo{volume}{100}},
  \bibinfo{pages}{017203} (\bibinfo{year}{2008}).

\bibitem[{\citenamefont{Kuklov et~al.}(2008)\citenamefont{Kuklov, Matsumoto,
  Prokof'ev, Svistunov, and Troyer}}]{Kuklov2008}
\bibinfo{author}{\bibfnamefont{A.~B.} \bibnamefont{Kuklov}},
  \bibinfo{author}{\bibfnamefont{M.}~\bibnamefont{Matsumoto}},
  \bibinfo{author}{\bibfnamefont{N.~V.} \bibnamefont{Prokof'ev}},
  \bibinfo{author}{\bibfnamefont{B.~V.} \bibnamefont{Svistunov}},
  \bibnamefont{and} \bibinfo{author}{\bibfnamefont{M.}~\bibnamefont{Troyer}},
  \bibinfo{journal}{Phys. Rev. Lett.} \textbf{\bibinfo{volume}{101}},
  \bibinfo{pages}{050405} (\bibinfo{year}{2008}).

\bibitem[{\citenamefont{Jiang et~al.}(2008)\citenamefont{Jiang, Nyfeler,
  Chandrasekharan, and Wiese}}]{Jiang2008}
\bibinfo{author}{\bibfnamefont{F.~J.} \bibnamefont{Jiang}},
  \bibinfo{author}{\bibfnamefont{M.}~\bibnamefont{Nyfeler}},
  \bibinfo{author}{\bibfnamefont{S.}~\bibnamefont{Chandrasekharan}},
  \bibnamefont{and} \bibinfo{author}{\bibfnamefont{U.~J.} \bibnamefont{Wiese}},
  \bibinfo{journal}{J. Stat. Mech. - Th. and Expt.}
  \textbf{\bibinfo{volume}{P02009}} (\bibinfo{year}{2008}).

\bibitem[{\citenamefont{Sirker et~al.}(2006)\citenamefont{Sirker, Weihong,
  Sushkov, and Oitmaa}}]{Sirker2006}
\bibinfo{author}{\bibfnamefont{J.}~\bibnamefont{Sirker}},
  \bibinfo{author}{\bibfnamefont{W.}~\bibnamefont{Zheng}},
  \bibinfo{author}{\bibfnamefont{O.~P.} \bibnamefont{Sushkov}},
  \bibnamefont{and} \bibinfo{author}{\bibfnamefont{J.}~\bibnamefont{Oitmaa}},
  \bibinfo{journal}{Phys. Rev. B.} \textbf{\bibinfo{volume}{73}},
  \bibinfo{pages}{184420} (\bibinfo{year}{2006}).

\bibitem[{\citenamefont{Darradi et~al.}(2008)\citenamefont{Darradi, Derzhko,
  Zinke, Schulenburg, Kruger, and Ricther}}]{Darradi2008}
\bibinfo{author}{\bibfnamefont{R.}~\bibnamefont{Darradi}},
  \bibinfo{author}{\bibfnamefont{O.}~\bibnamefont{Derzhko}},
  \bibinfo{author}{\bibfnamefont{R.}~\bibnamefont{Zinke}},
  \bibinfo{author}{\bibfnamefont{J.}~\bibnamefont{Schulenburg}},
  \bibinfo{author}{\bibfnamefont{S.~E.} \bibnamefont{Kruger}},
  \bibnamefont{and} \bibinfo{author}{\bibfnamefont{J.}~\bibnamefont{Ricther}},
  \bibinfo{journal}{Phys. Rev. B.} \textbf{\bibinfo{volume}{78}},
  \bibinfo{pages}{214412} (\bibinfo{year}{2008}).

\bibitem[{\citenamefont{Gelfand and Singh}(2000)}]{Gelfand2000}
\bibinfo{author}{\bibfnamefont{M.~P.} \bibnamefont{Gelfand}} \bibnamefont{and}
  \bibinfo{author}{\bibfnamefont{R.~R.~P.} \bibnamefont{Singh}},
  \bibinfo{journal}{Adv. in Phys.} \textbf{\bibinfo{volume}{49}},
  \bibinfo{pages}{93} (\bibinfo{year}{2000}).

\bibitem[{\citenamefont{Singh}(1989)}]{Singh1989}
\bibinfo{author}{\bibfnamefont{R.~R.~P.} \bibnamefont{Singh}},
  \bibinfo{journal}{Phys. Rev. B} \textbf{\bibinfo{volume}{39}},
  \bibinfo{pages}{9760} (\bibinfo{year}{1989}).

\end{thebibliography}

\end{document}